\documentclass[12pt]{article}

\title{Quadratic Poisson algebras for two dimensional classical
superintegrable systems and quadratic associative algebras for
quantum superintegrable systems. }

\author{C. Daskaloyannis\thanks{e:mail address: daskalo@auth.gr}\\
                  {\it  Physics Department,}\\
        {\it Aristotle University of Thessaloniki,}\\
                {\it  54006 Thessaloniki, Greece}
        }

\date{February 2000}
\begin{document}

 \maketitle

\begin{abstract}
The integrals of motion of the classical two dimensional
superintegrable systems with quadratic integrals of motion  close
in a restrained quadratic Poisson algebra, whose the general form
is investigated. Each classical superintegrable problem has a
quantum counterpart, a quantum superintegrable system. The
quadratic Poisson algebra is deformed to a quantum associative
algebra, the finite dimensional representations of this algebra
are calculated by using a deformed parafermion oscillator
technique. It is shown that, the finite dimensional
representations of the quadratic algebra are determined by the
energy eigenvalues of the superintegrable system. The calculation
of energy eigenvalues is reduced to the solution of algebraic
equations, which are universal for all two dimensional
superintegrable systems with quadratic integrals of motion.
\end{abstract}

\vfill

Running title: Quadratic algebras for superintegrable systems


\newpage

\section{Introduction}\label{sec:Alg}
In classical mechanics, integrable system is a system possessing
more constants of motion in addition to the energy. A
comprehensive review of the two-dimensional integrable classical
systems  is given by Hietarinta \cite{Hietarinta87}, where the
space was assumed to be flat. The case of non flat space is under
current investigation \cite{Ranada97,RaSan99}. An interesting
subset of the totality of integrable systems is the set of
systems, which possess a maximum number of integrals, these
systems are termed as superintegrable ones. The Coulomb and the
harmonic oscillator potentials are the most familiar classical
superintegrable systems, whose their quantum counterpart has nice
symmetry properties, which are described by the $so(N+1)$ and
$su(N)$ Lie algebras respectively.

The Hamiltonian of a classical system is generally a quadratic
function of the momenta. In the case of the flat space, all the
known two dimensional superintegrable systems with quadratic
integrals of motion are simultaneously separable in more than two
orthogonal coordinate systems \cite{Fris}. The integrals of motion
of a two dimensional superintegrable system in flat space close in
a restrained classical  Poisson algebra
\cite{BoDasKo93,BoDasKo94,KaMiPogo96}. The study of the quadratic
Poisson algebras is a matter under investigation, related to
several branches of physics as: the solution of the classical Yang
- Baxter equation \cite{Sklyanin83},  the two dimensional
superintegrable systems in flat space or on the sphere
 \cite{KaMiPogo96}, the statistics \cite{EssRitt94} or  the case
 of "exactly solvable" classical problems \cite{GLZ-1992}.

 The quantization of a classical integrable system corresponds
generally  to a quantum integrable system. The mechanism of
quantum deformation of a classical system to a quantum one is not
fully understood. Initially the problem of quantization of
classical superintegrable system was viewed as a relatively simple
and somehow trivial problem \cite{Korsch82}, but several authors
have proved that this quantization procedure has  to add
correction terms to the integrals of motion or to the Hamiltonian,
these correction terms are  of  order ${\cal O}(\hbar^2)$
\cite{HietGramm89,Hietarinta98}. The result of the quantum
deformation of a superintegrable system is realized by the shift
of the classical Poisson algebra  to some quantum polynomial
associative algebra. The same fact is true in the case of
quadratic Poisson algebra corresponding to the Yang - Baxter
equation \cite{Sklyanin83}, which is turned to a quantum quadratic
associative algebra \cite{Sklyanin84} with four generators. The
same idea was discussed in reference \cite{GLZ-1992}, where the
classical problems, which are expressed by a quadratic Poisson
algebra are mapped to quantum ones described by the corresponding
quantum operator quadratic algebra. The same shift is indeed true
for the superintegrable systems, where the classical ones
correspond to the quantum ones and the classical quadratic Poisson
algebra is mapped to a quadratic associative
algebra\cite{GLZ-R1}--\cite{KaMiHaPo99}.

In this paper we show that the deformation of the classical
Poisson algebra to a quadratic associative algebra implies a
deformation of the parameters of the quadratic algebra.   The
general form of the quadratic algebras, which are encountered in
the case of the two dimensional quantum superintegrable systems,
is investigated.  In references
\cite{BoDasKo93,BoDasKo94,GLZ-R1,GLZ-R2,GLZ-R3,Higgs79,LeVin95,GGZ91,Zhedanov92}
was conjectured that, the energy eigenvalues correspond to finite
dimensional representations of the latent quadratic algebras.
Granovskii et al in \cite{GLZ-1992} studied the representations of
the quadratic Askey - Wilson algebras $QAW(3)$. Using there the
proposed ladder representation, the finite dimensional
representations are calculated and this method was applied to
several superintegrable systems
\cite{GLZ-R1,GLZ-R2,GLZ-R3,LeVin95,Zhedanov92}. Another method
\cite{BoDasKo93,BoDasKo94} for calculating the finite dimensional
representations is the use of the deformed oscillator algebra
\cite{Das1} and their finite dimensional version which are termed
as generalized deformed parafermionic algebras\cite{Quesne}. The
main task of this paper is to reduce the calculations of
eigenvalues to a system  of two algebraic equations with two
parameters to be determined. These equations are universal
equations, which are valid of all superintegrable systems, with
quadratic integrals of motion.

This paper is organized as follows: In section \ref{sec:Poisson}
the general form of the quadratic Poisson algebra for a two
dimensional system with quadratic integrals of motion is derived.
In section \ref{sec:PoissonSuper} the special form of the Poisson
algebra of the known two dimensional superintegrable systems in
flat space is written. In section \ref{sec:AssocQuadrAlgebra} the
quantum version of the Poisson quadratic algebra is studied. The
deformed parafermionic oscillator algebra is reviewed and the
oscillator realization of the quadratic algebras is realized. The
finite dimensional representations of the quadratic algebras are
generated by using the technique of deformed parafermionic
algebras. The problem is reduced to the solution of a system of
two algebraic equations in section \ref{sec:Representations}. In
section \ref{sec:QuantumQuadratic} the energy eigenvalues of all
the known superintegrable systems in the flat two dimensional
space  are determined by solving the appropriate algebraic
equations. Finally in section \ref{sec:Discussion} there is a
discussion  of the results of this paper.

\section{Quadratic Poisson Algebras}\label{sec:Poisson}
Let consider a two dimensional superintegrable system. The general
form of the Hamiltonian is:
\begin{equation}
H= a(q_1,q_2)p_1^2+2 b(q_1,q_2) p_1 p_2 + c(q_1,q_2) p_2^2 +
V(q_1,q_2) \label{eq:ClassicalHamiltonian}
\end{equation}
this Hamiltonian is a quadratic form of the momenta. The system is
superintegrable, therefore there are two additional integrals of
motion $A$ and $B$. In that section, we consider that, these
integrals of motion are quadratic functions of the momenta, i.e.
they are given by the general forms:
\[
\begin{array}{rl}
A=& A(q_1,q_2,p_1,p_2) = c(q_1,q_2)p_1^2+2 d(q_1,q_2) p_1 p_2 +
e(q_1,q_2) p_2^2 +\\ &+ f(q_1,q_2)p_1 +g(q_1,q_2)p_2 +Q(q_1,q_2)
\end{array}
\]
The integral of motion $B$ is assumed to be indeed a quadratic
form, which is analogous to above one.
\[
\begin{array}{rl}
B=& B(q_1,q_2,p_1,p_2) = h(q_1,1_2)p_1^2+2 k(q_1,q_2) p_1 p_2 +
l(q_1,q_2) p_2^2 +\\ &+ m(q_1,q_2)p_1 +n(q_1,q_2)p_2 + S(q_1,q_2)
\end{array}
\]
 By definition the following
relations are satisfied:
\begin{equation}
\left\{ H, A\right\}_{P}= \left\{ H, B\right\}_{P}=0
\label{eq:PBIntegrals}
\end{equation}
where $\left\{ \;.\;,\;.\; \right\}_{P}$ is the usual Poisson
bracket.

From the integrals of motion $A,B$, we can construct the integral
of motion:
\begin{equation}
C=\left\{ A, B\right\}_{P}
 \label{eq:ClassicalC} \end{equation}
The integral of motion $C$ is not a new independent integral of
motion, which is a cubic function of the momenta. The integral $C$
is not independent from the integrals $H,\;A$ and $B$ as it will
be shown later. The fact that, the integral $C$ is a cubic
function of momenta, implies the impossibility of expressing $C$
as a polynomial function of the other integrals, which are
quadratic functions of momenta.
 Starting from the integral of motion $C$, we can
construct the (non independent) integrals $\left\{ A,
C\right\}_{P}$ and $\left\{ B, C\right\}_{P}$. These integrals are
quartic functions of the momenta, i.e. functions of fourth order.
 Therefore these integrals could be expressed as quadratic combinations of the
integrals $H,\;A$, and $B$. Therefore the following relations are
assumed to be valid:
\begin{equation}
\left\{ A, C\right\}_{P}= \alpha A^2 + \beta B^2 + 2 \gamma A B +
\delta A + \epsilon B +\zeta \label{eq:AC}
\end{equation}
and
\begin{equation}
 \left\{ B , C \right\}_P = a A^2 + b B ^2 + 2 c A B
+ d A + e B + z \label{eq:BC}
\end{equation}
We can take appropriate a linear combination  of the integrals $A$
and $B$ and
 we can always consider the case
$\beta =0$.

The Jacobi equality for the Poisson brackets induces the relation
\[
 \left\{ A, \left\{ B, C \right\}_P \right\}_P =
  \left\{ B, \left\{ A, C
\right\}_P \right\}_P
\]
 The following  relations
 $$ b = - \gamma , \quad c= - \alpha
\quad \mbox{and} \quad e= - \delta $$
 must be satisfied.

 The integrals $A,\;B$ and $C$ satisfy the  quadratic Poisson
 algebra:
\begin{equation}
\begin{array}{rcl}
\left\{ A, B\right\}_{P}&=&C \\
 \left\{ A , C \right\}_P &=& \alpha A^2
+2 \gamma  A, B  + \delta A + \epsilon B + \zeta\\
 \left\{ B , C \right\}_P &=& a A^2 - \gamma B ^2 - 2\alpha  A B + d A
-\delta B + z
\end{array}
\label{eq:PoissonAlgebra}
\end{equation}
where $\alpha,\;\gamma,\;a$ are constants and
\[
\begin{array}{l}
\delta=\delta(H)= \delta_0 + \delta_1 H \\
 \epsilon=\epsilon(H)=\epsilon_0 + \epsilon_1 H\\
\zeta= \zeta(H) = \zeta_0 + \zeta_1 H + \zeta_2 H^2\\ d=d(H)=
d_0+d_1 H \\ z= z(H) = z_0 + z_1 H + z_2 H^2
\end{array}
\]
where $\delta_i,\;\epsilon_i,\;\zeta_i,\;d_i$ and $z_i$ are
constants. The associative algebra, whose the generators satisfy
equations (\ref{eq:PoissonAlgebra}), is the general form of the
closed Poisson algebra of the integrals of superintegrable systems
with integrals quadratic in momenta.

The quadratic Poisson algebra (\ref{eq:PoissonAlgebra}) possess a
Casimir which is a  function of momenta of degree 6 and it is
given by:
\begin{equation}
\begin{array}{rl}
 K=& C^2-2\alpha A^2 B - 2 \gamma A B^2 -2 \delta A B-\\
 &- \epsilon B^2 -2 \zeta B+\frac{2}{3}a A^3 + d A^2 +2 z A=\\
 =& k_0 + k_1 H + k_2 H^2+k_3 H^3
\end{array}
 \label{eq:ClassicalCasimir}
\end{equation}
Obviously
\[
\left\{ K,A \right\}_P = \left\{ K,B \right\}_P =\left\{ K,C
\right\}_P =0
\]
Therefore the integrals of motion of a superintegrable two
dimensional system with quadratic integrals of motion close a
constrained classical quadratic Poisson algebra
(\ref{eq:PoissonAlgebra}), corresponding to a Casimir equal at
most to a cubic function of the Hamiltonian
(\ref{eq:ClassicalCasimir}).

In the general case of a superintegrable system the integrals are
not necessarily quadratic functions of the momenta, but rather
polynomial functions of the momenta. The case of the systems with
a quadratic and a cubic integral of motion are recently studied by
Tsiganov \cite{Tsiganov00}.
 The general form of the Poisson
algebra of generators $A,\; B,$ and $C$ is characterized  by  a
polynomial function $h(A,B)$, which satisfy  the following
equations:
\begin{equation}
\begin{array}{rcl}
\left\{ A, B\right\}_{P}&=&C \\
 \left\{ A , C \right\}_P &=&{\partial h}/{\partial B}\\
 \left\{ B , C \right\}_P &=&-{\partial h}/{\partial A}
\end{array}
\label{eq:GeneralPoissonAlgebra}
\end{equation}
and the Casimir of the algebra is given by
\begin{equation}
K=K(H) =  C^2 - 2 h(A,B), \qquad \left\{ K, A\right\}_{P}=\left\{
K, B\right\}_{P}=0 \label{eq:GenetalPoissonCasimir}
\end{equation}
where $h(A,B)$ is a polynomial function of the integrals of motion
$A$ and $B$. In the case of the quadratic Poisson algebra
(\ref{eq:PoissonAlgebra}) the form of the function $h(A,B)$ is
given by equation (\ref{eq:ClassicalCasimir}):
\[
\begin{array}{rl}
h(A,B)=& -\frac{a}{3}A^3 + \alpha A^2B + \gamma A B^2\\
&-\frac{d}{2} A^2 + \delta A B + \frac{\epsilon}{2}B^2\\ &-z A +
\zeta B
 \end{array}
\]

In the general case of a two dimensional superintegrable system,
with quadratic Hamiltonian,  one integral $A$ of order $m$ in
momenta and one integral $B$ of order $n$ ($n\ge m$), the general
form of the function $h(A,B)$ can be given by the general form:
\[
h(A,B) = h_0(A) + h_1(A) B + h_2(A) B^2
\]
where $h_i(A)$ are polynomials of the integrals $A$ and $H$. The
proof of this assumption is based on the dependence of the
integrals of motion on the momenta. For simplicity reasons, the
proof of this proposition will not be given here.

\section{Poisson algebras for superintegrable
systems}\label{sec:PoissonSuper}

 Let consider the superintegrable
systems with quadratic integrals of motion, these potentials are
given by several authors starting from different but comparable
points of view. In references \cite{Hietarinta87,Ranada97} the
integrals of motion are generated by solving the Darboux
conditions for integrability of quaratic integrals. In
\cite{Fris}the Hamilton - Jacobi equation is solved by separation
of variables nad the two dimensional Hamiltonians which are
separable in more than one coordinate system are obtained. The
separation of variables is essential for solving the quantum
counterpart of the superintegrable system and the solution of the
associate Schr\"{o}dinger equations is given in \cite{KaMiPogo96}.
Using this method the quantum superintegrable systems have been
solved on the sphere\cite{KaMiPogo96} and the
hyperboloid\cite{KaMiPogo96,KaMiHaPo99}. From classical point of
view the super integrable are given in \cite{RaSan99}, while the
case of a pseudo Euclidean kinetic term has been studied in
\cite{Ranada97}. The extension on the systems with a quadratic and
a cubic integral of motion is sytematized in \cite{Tsiganov00}.

 In this section we consider the
case superintegrable systems gin in ref \cite{KaMiPogo96}, because
in the next sections we study the quantum counterparts of these
potentials. In this paper the following superintegrable systems
are considered:

\noindent  {\bf Potential i):}
 \[\displaystyle H=
\frac{1}{2}\left(p_x^2+p_y^2+ \omega^2 r^2+
\frac{\mu_1}{x^2}+\frac{\mu_2}{y^2}\right)
\]
This potential has the following independent integrals of motion:
\[
A=p_x^2+\omega^2 x^2+\frac{\mu_1}{x^2}
\]
and
\[
 B=\left(x p_y - y
p_x\right)^2+r^2\left(\frac{\mu_1}{x^2}+\frac{\mu_2}{y^2}\right)
\]
The constants, which characterize  the corresponding quadratic
algebra (\ref{eq:PoissonAlgebra}), are given by:
\[
\begin{array}{lll}
\alpha=-8,& \gamma=0,& \delta=16 H,\\
 \epsilon=-16 \omega^2,&
\zeta=16(\mu_1+\mu_2)\omega^2\\ a=0,& d=0, &
z=16(\mu_2-\mu_1)\omega^2
\end{array}
\]
the value of the Casimir (\ref{eq:ClassicalCasimir}) is:
\[
K= -16\left((\mu_2-\mu_1)^2\omega^2+4 \mu_1 H^2 \right)
\]

\noindent  {\bf Potential ii):}
\[
\displaystyle H =\frac{1}{2}\left(p_x^2+p_y^2+
  \omega^2\left(4x^2+y^2\right)+
\frac{\mu}{y^2}\right)
\]
 This potential has the following independent
integrals of motion:
\[
A=p_x^2+4\omega^2 x^2
\]
 and
\[
 B=\left(x p_y -y p_x\right)p_y+ \frac{\mu x}{y^2} -\omega^2 x y^2
\]
The constants, which characterize  the corresponding quadratic
algebra (\ref{eq:PoissonAlgebra}), are given by:
\[
\begin{array}{lll}
\alpha=0,&\gamma=0,& \delta=0,\\
 \epsilon=-16 \omega^2,&
\zeta=0,\\ a=-6, & d=16 H,&z=8 \mu_2\omega^2-8 H^2
\end{array}
\]
the value of the Casimir (\ref{eq:ClassicalCasimir}) is:
\[
K=0
\]

\noindent  {\bf Potential iii):}
\[
 H =\frac{1}{2}\left(p_x^2+p_y^2
  +\frac{k}{r} + \frac{1}{r}\left(\frac{\mu_1}{r+x}+\frac{\mu_2}{r-x}\right) \right)
\]
This potential has the following independent integrals of motion:
\[
 A=(xp_y-yp_x)^2+r \left(\frac{\mu_1}{r+x}+\frac{\mu_2}{r-x}\right)
\]
and
\[
B=\left(x p_y - y p_x\right)p_y -\frac{\mu_1}{2
r}\frac{r-x}{r+x}+\frac{\mu_2}{2 r}\frac{r+x}{r-x}+\frac{k x}{2r}
\]
The constants, which characterize  the corresponding quadratic
algebra (\ref{eq:PoissonAlgebra}), are given by:
\[
\begin{array}{lll}
\alpha=0,&\gamma=-2,& \delta=0,\\
 \epsilon=0,& \zeta=k
(\mu_1-\mu_2), \\a=0,& d=-8 H,&z=4(\mu_1+ \mu_2)H- k^2/2
\end{array}
\]
the value of the Casimir (\ref{eq:ClassicalCasimir}) is:
\[
K=2(\mu_1- \mu_2)^2 H -k^2(\mu_1+ \mu_2)
\]

\noindent  {\bf Potential iv):}
\[
 H =\frac{1}{2}\left(p_x^2+p_y^2
  +\frac{k}{r} +\mu_1
  \frac{\sqrt{r+x}}{r}+\mu_2\frac{\sqrt{r-x}}{r
  } \right)
\]
This potential has the following independent integrals of motion:
\[
A=(yp_x-xp_y) p_y+ \frac{\mu_1 (r-x)\sqrt{r+u}}{2r} -\frac{\mu_2
(r+x)\sqrt{r-u}}{2r}-\frac{k x}{2 r}
\]
and
\[
B=\left(x p_y - y p_x\right)p_x- \frac{\mu_1 x\sqrt{r-u}}{2r}
+\frac{\mu_2 x\sqrt{r+u}}{2r}-\frac{k y}{2 r}
\]
The constants, which characterize  the corresponding quadratic
algebra (\ref{eq:PoissonAlgebra}), are given by:
\[
\begin{array}{lll}
\alpha=0,& \gamma=0,& \delta=0,\\
 \epsilon=2 H,& \zeta=-\mu_1
\mu_2/2 ,\\ a=0,& d=-2 H,& z=\frac{\mu_1^2-\mu_2^2}{4}
\end{array}
\]
the value of the Casimir (\ref{eq:ClassicalCasimir}) is:
\[
K=- {k^2}H/{2}  -k(\mu_1^2+\mu_2^2)/{4}
\]

\section{The quadratic associative
algebra}\label{sec:AssocQuadrAlgebra}

 The quantum counterparts of
the  classical systems, which have been studied in section
\ref{sec:Poisson}, are quantum superintegrable systems. The
quadratic classical Poisson algebra (\ref{eq:PoissonAlgebra})
possesses a quantum counterpart, which is a quadratic associative
algebra of operators.  The form of the quadratic algebra is
similar to the classical Poisson algebra, the involved constants
are generally functions of $\hbar$ and they should coincide with
the classical constants in the case $\hbar\to 0$.
 Let consider the quadratic associative algebra generated
by the generators $\left\{ A,\,  B,\,  C\right\}$,
 which satisfy the commutation relations
\begin{equation}
\begin{array}{l}
\left[ A, B \right] = C\\ \left[ A , C \right] = \alpha A^2 +
\beta B^2 + \gamma \left\{ A, B \right\} + \delta A + \epsilon B +
\zeta \\ \left[ B , C \right] = a A^2 + b B ^2 + c\left\{ A, B
\right\} + d A + e B + z
\end{array}
\label{eq:prealgebra}
\end{equation}
After rotating the generators $A$ and $B$,
 we can always consider the case
$\beta =0$.

 The Jacobi equality for the commutator induces the relation
$$ \left[ A, \left[ B, C \right] \right] = \left[ B, \left[ A, C
\right] \right] $$
 the following  relations
 $$ b = - \gamma , \quad c= - \alpha
\quad \mbox{and} \quad e= - \delta $$
 must be satisfied, and consequently the
general form of the quadratic algebra (\ref{eq:prealgebra}) can be
explicitly written as follows:
\begin{equation}
\left[ A, B \right] = C \label{eq:algebra1}
\end{equation}
\begin{equation}
\left[ A , C \right] = \alpha A^2  + \gamma \left\{ A, B \right\}
+ \delta A + \epsilon B + \zeta \label{eq:algebra2}
\end{equation}
\begin{equation}
\left[ B , C \right] = a A^2 - \gamma B ^2 - \alpha \left\{ A, B
\right\} + d A -\delta B + z \label{eq:algebra3}
\end{equation}
The Casimir of this algebra is given by:
\begin{equation}
\begin{array}{rl}
K =& C^2 - \alpha \left\{ A^2, B \right\} -\gamma \left\{ A, B^2
\right\} + ( \alpha \gamma - \delta ) \left\{ A, B \right\}+\\ +&
(\gamma^2 - \epsilon) B^2 + ( \gamma \delta - 2 \zeta ) B +\\ +&
\frac{2a}{3} A^3 + ( d + \frac{a \gamma}{3} + \alpha^2) A^2 + (
\frac{a \epsilon}{3} + \alpha \delta + 2 z ) A
\end{array}
\label{eq:Casimir}
\end{equation}
another useful form of the Casimir of the algebra is given by:
\begin{equation}
\begin{array}{rl}
K=& C^2 +\frac{2 a}{3} A^3 -\frac{ \alpha}{3} \left\{A,A,B\right\}
- \frac{ \gamma}{3} \left\{A,B,B\right\} +\\ &+ \left(
{\frac{2{\alpha^2}}{3}} + d + {\frac{2a\gamma}{3}} \right)A^2   +
\left( -\epsilon + \frac{2\gamma^2}{3} \right) B^2  +\\
&+\left(-\delta + \frac{a \gamma}{3}\right)\{ A, B \} + \left(
\frac{2\alpha\delta}{3} + \frac{a\epsilon}{3} +
  \frac{d\gamma}{3} + 2z
\right)A+\\ &+ \left( -\frac{\alpha\epsilon}{3} +
  \frac{2\delta \gamma}{3} - 2\zeta
\right) B + \frac{\gamma z}{3} - \frac{\alpha\zeta}{3}
\end{array}
\label{eq:Casimir1}
\end{equation}
where
\[
\left\{A,B,C\right\}=ABC+ACB+BAC+BCA+CAB+CBA
\]
This quadratic algebra has many similarities to the Racah algebra
$QR(3)$, which is a special case of the Askey - Wilson algebra
$QAW(3)$. The algebra (\ref{eq:algebra1} --  \ref{eq:algebra3})
does not coincide with the Racah algebra $QR(3)$, if $a \ne 0$ in
the relation (\ref{eq:algebra3}).
 Unless this difference between  (\ref{eq:prealgebra}) and $QR(3)$ algebra
  a representation theory can be constructed by following the same procedures
   as they were described by Granovskii, Lutzenko and Zhedanov in ref.
\cite{GLZ-1992,GLZ-R1,GLZ-R2}. In this paper we shall give a
realization of this algebra using the deformed oscillator
techniques\cite{Das1}. The finite dimensional representations of
the algebra  (\ref{eq:prealgebra}) will be constructed by
constructing a realization of the algebra  with
 the generalized parafermionic algebra introduced by Quesne\cite{Quesne}.

\section{Deformed Parafermionic Algebra}\label{sec:Para}

Let now consider a realization of the algebra (\ref{eq:algebra1}
--  \ref{eq:algebra3}), by using
 of the deformed oscillator   technique, i.e. by using
 a deformed
oscillator algebra\cite{Das1} $\left\{ b^\dagger, b, {\cal N}
\right\}$, which satisfies the
\begin{equation}
\left[ {\cal N}, b^\dagger \right] = b^\dagger, \quad \left[ {\cal
N}, b \right] = -b, \quad b^\dagger b = \Phi\left({\cal N}\right),
\quad b b^\dagger = \Phi \left({\cal N}+1\right)
\label{eq:DefOsci}
\end{equation}
where the function $\Phi(x)$ is a "well behaved" real function
which satisfies the boundary condition:
\begin{equation}
\Phi(0)=0, \quad \mbox{and} \quad  \Phi(x)>0  \quad \mbox{for}
\quad x>0
 \label{eq:restriction1}
\end{equation}
 As it is well known\cite{Das1} this
constraint imposes the existence a Fock type representation of the
deformed oscillator algebra, which is bounded by bellow, i.e.
there is a Fock basis $|n>,\; n=0,1,\ldots$ such that
\begin{equation}
\begin{array}{l}
{\cal N}|n>=n|n>\\ b^\dagger | n> = \sqrt{ \Phi \left(n+1\right)}
| n+1>,\quad  n=0,1,\ldots\\ b|0>=0\\ b|n>= \sqrt{ \Phi
\left(n\right)} | n-1>,\quad  n=1,2,\ldots
\end{array}
\label{eq:Fock}
\end{equation}
The Fock representation (\ref{eq:Fock}) is bounded by bellow. The
generalized deformed algebra given in ref \cite{Das1} is
equivalent to several deformed oscillator schemes as the Odaka-
Kishi - Kamefuchi unification scheme \cite{OKK91}, the Beckers-
Debergh unification scheme \cite{BeDe91}, The Fibonacci oscillator
\cite{Arik92}, for a discussion of deformation schemes see
\cite{BD93}

In the case of nilpotent deformed oscillator algebras, there is a
positive integer $p$, such that $$ b^{p+1}=0, \quad
\left(b^\dagger\right)^{p+1}=0 $$ the above equations imply that
\begin{equation}
\Phi(p+1)=0,
 \label{eq:restriction2}
\end{equation}
In that case the deformed oscillator (\ref{eq:DefOsci}) has a
finite dimensional representation, with dimension equal to $p+1$,
this kind of oscillators are called deformed parafermion
oscillators of order $p$.

An interesting property of the deformed parafermionic algebra is
that the existence of a faithful finite dimensional representation
of the algebra implies that:
\begin{equation}
 {\cal N} \left(  {\cal N} -1 \right) \left(  {\cal N} -2 \right)\cdots \left(  {\cal N} -p \right) = 0
\label{eq:Nrestriction}
\end{equation}
The above restriction and the constraints (\ref {eq:restriction1})
and (\ref {eq:restriction2}) imply that the general form of the
structure function $\Phi( {\cal N} )$ has the general
form\cite{Quesne}:
 \[ \Phi( {\cal N} ) = {\cal N}( p+1-{\cal N}) (
a_0 + a_1 {\cal N}+ a_2 {\cal N}^2 +\cdots a_{p-1} {\cal N}^{p-1}
) \] A systematic study and applications of the parafermionic
oscillator is given in references
\cite{Quesne,BeDeQue96,KliPly99,Debergh95}.

We shall show, that there is a realization of the quadratic
algebra , such that
\begin{eqnarray}
A= A\left({\cal N}\right)\label{eq:A}\\ B=b\left({\cal N}\right)+
b^\dagger \rho \left({\cal N}\right)+ \rho \left({\cal N}\right)b
\label{eq:B}
\end{eqnarray}
where the $A[x],\; b[x]$ and $\rho(x)$ are functions, which will
be determined. In that case (\ref{eq:algebra1}) implies:
\begin{equation}
C=\left[A,B\right]\; \Rightarrow\; C=b^\dagger \Delta A\left({\cal
N}\right)\rho \left({\cal N}\right) -\rho \left({\cal
N}\right)\Delta A\left({\cal N}\right) b \label{eq:C}
\end{equation}
where
 $$ \Delta A\left({\cal N}\right) = A\left({\cal N}+1\right)
- A\left({\cal N}\right) $$
 Using equations (\ref{eq:A}),
(\ref{eq:B}) and (\ref{eq:algebra2}) we find:
\begin{equation}
\begin{array}{rl}
[A,C]=& [ A\left({\cal N}\right),
 b^\dagger \Delta A\left({\cal N}\right)\rho\left({\cal N}\right)-
\rho\left({\cal N}\right)\Delta A\left({\cal N}\right) b ]=\\
 =& b^\dagger \left( \Delta A\left({\cal N}\right)\right)^2 \rho\left({\cal N}\right)+
 \rho\left({\cal N}\right)\left( \Delta A\left({\cal N}\right)\right)^2 b=\\
 =&
 \alpha A^2  + \gamma \left\{ A, B \right\} + \delta A + \epsilon B + \zeta=\\
=& b^\dagger \left( \gamma \left(  A\left({\cal N}+1\right)+
A\left({\cal N}\right)\right)+\epsilon \right)\rho\left({\cal
N}\right)+
\\
&+ \rho\left({\cal N}\right) \left( \gamma \left(  A\left({\cal
N}+1\right)+ A\left({\cal N}\right)\right)+ \epsilon \right)b + \\
&+
 \alpha A\left({\cal N}\right)^2  + 2\gamma  A\left({\cal N}\right) b\left({\cal N}\right)
  + \delta A\left({\cal N}\right) + \epsilon B\left({\cal N}\right) + \zeta
\end{array}
\label{eq:detail1}
\end{equation}
therefore we have the following relations:
\begin{eqnarray}
 \left( \Delta   A\left({\cal N}\right) \right)^2=
 \gamma \left(  A\left({\cal N}+1\right)+ A\left({\cal N}\right)\right)+
 \epsilon
 \label{eq:eqn1}\\
 \alpha A\left({\cal N}\right)^2  + 2\gamma  A\left({\cal N}\right) b\left({\cal N}\right)
  + \delta A\left({\cal N}\right) + \epsilon B\left({\cal N}\right) + \zeta =0
  \label{eq:eqn2}
\end{eqnarray}
while the function $\rho\left({\cal N}\right)$ can be arbitrarily
determined. In fact  this function can be fixed, in order to have
a polynomial structure function $\Phi(x)$ for the deformed
oscillator algebra (\ref{eq:DefOsci}).

The solutions of equation (\ref{eq:eqn1})  depend on the value of
the parameter $\gamma$, while the function $b({\cal N})$ is
uniquely determined by equation (\ref{eq:eqn2}) (provided that
almost one among the parameters $\gamma$ or $\epsilon$ is not
zero).
  At this stage,
 the cases $\gamma \ne 0$ or  $\gamma =0$, should be treated separately.
 We can see that:
\begin{itemize}
\item[{\bf Case 1}:]  $\gamma \ne 0$\\
In that case the solutions of equations (\ref{eq:eqn1}) and
(\ref{eq:eqn2}) are given by:
\begin{equation}
 A\left({\cal N}\right) =
 \frac{\gamma}{2} \left(
 ({\cal N}+u)^2-1/4
-\frac{\epsilon}{ \gamma^2} \right) \label{eq:sol1}
\end{equation}
\begin{equation}
\begin{array}{rl}
b \left({\cal N}\right)=& -{\frac{ \alpha  \left( ({\cal
N}+u)^2-1/4 \right) }{4}} + {\frac{\alpha\,\epsilon -
     \delta\,\gamma}{2\,
     {{\gamma}^2}}}
-
\\
&-{\frac{ \alpha\,{{\epsilon}^2}
           - 2\,\delta\,\epsilon\,
      \gamma + 4\,{{\gamma}^2}\,
      \zeta}{4\,{{\gamma}^4}}}
\frac{1}{  \left( ({\cal N}+u)^2-1/4 \right) }
\end{array}
\label{eq:sol2}
\end{equation}

\item[{\bf Case 2}:]  $\gamma = 0, \; \epsilon \ne 0 $\\
 The solutions of equations (\ref{eq:eqn1}) and (\ref{eq:eqn2}) are given by:
\begin{equation}
A({\cal N}) = \sqrt{\epsilon}  \left( {\cal N}+u \right)
\label{eq:sol1a}
\end{equation}
\begin{equation}
b({\cal N}) =-\alpha \left( {\cal N}+u \right)^2- \frac{\delta}{
\sqrt{\epsilon} }   \left( {\cal N}+u \right)
  -
\frac{\zeta}{\epsilon} \label{eq:sol2a}
\end{equation}
\end{itemize}

The constant $u$ will be determined later.

Using the above definitions of equations $A({\cal N})$ and
$b({\cal N})$, the left hand side and right hand side of equation
(\ref{eq:algebra3}) gives the following equation:
\begin{equation}
\begin{array}{l}
2\,\Phi({\cal N}+1)\left( \Delta A\left({\cal N}\right)
+\frac{\gamma}{2} \right)  \rho({\cal N})
-
2\,\Phi({\cal N})\left( \Delta A\left({\cal N}-1\right)
-\frac{\gamma}{2} \right) \rho({\cal N}-1) =\\ =a A^2\left({\cal
N}\right) -\gamma b^2({\cal N}) -2 \alpha A\left({\cal N}\right)
b({\cal N}) +d A\left({\cal N}\right)-\delta b({\cal N})+z
\end{array}
\label{eq:basic1}
\end{equation}
Equation (\ref{eq:Casimir}) gives the following relation:
\begin{equation}
\begin{array}{rl}
K=\\ =& \Phi({\cal N}+1)\left( \gamma^2 - \epsilon - 2 \gamma
A\left({\cal N}\right)  - \Delta A^2\left({\cal N}\right) \right)
\rho({\cal N})
 +\\
&+ \Phi({\cal N})\left( \gamma^2 - \epsilon - 2 \gamma
A\left({\cal N}\right)  - \Delta A^2\left({\cal N}-1\right)
\right) \rho({\cal N}-1) -\\ &-2 \alpha A^2\left({\cal N}\right)
b({\cal N}) +\left( \gamma^2 - \epsilon - 2 \gamma  A\left({\cal
N}\right) \right) b^2({\cal N})+\\ &+ 2 \left( \alpha \gamma -
\delta \right)  A\left({\cal N}\right) b({\cal N}) +\left( \gamma
\delta - 2\zeta \right) b({\cal N})+
\\
&+ \frac{2}{3}a A^3\left({\cal N}\right) + \left( d + \frac{1}{3}
a\gamma + \alpha^2 \right)A^2\left({\cal N}\right)+\\ &+\left(
\frac{1}{3} a \epsilon + \alpha \delta +2 z \right)A\left({\cal
N}\right)
\end{array}
\label{eq:basic2}
\end{equation}

Equations (\ref{eq:basic1}) and (\ref{eq:basic2}) are linear
functions of the expressions $\Phi\left({\cal N}\right)$ and
$\Phi\left({\cal N}+1\right)$, then
 the function
$\Phi\left({\cal N}\right)$ can be determined, if the function $
\rho({\cal N})$ is given. The solution of this system, i.e. the
function $\Phi\left({\cal N}\right)$ depends on two parameters $u$
and $K$ and it is given by the following formulae:
\begin{itemize}
\item[{\bf Case 1}:]  $\gamma \ne 0$\\
$$ \rho( {\cal N} ) = \frac{1}{3\cdot 2^{12}\cdot \gamma^8 ( {\cal
N}+u) ( 1 + {\cal N}+u ) ( 1 +2 ( {\cal N}+u) )^2} $$ and
\begin{equation}
\begin{array}{l}
\Phi({\cal N}) =-3072 \gamma^6 K (-1 + 2 ({\cal N}+u))^2-\\ - 48
\gamma^6
   (\alpha^2  \epsilon - \alpha  \delta  \gamma + a \epsilon  \gamma - d  \gamma^2) \cdot\\
\cdot (-3 + 2  ({\cal N}+u))  (-1 + 2  ({\cal N}+u))^4
   (1 + 2  ({\cal N}+u)) + \\
+\gamma^8  (3  \alpha^2 + 4  a  \gamma)  (-3 + 2  ({\cal N}+u))^2
(-1 + 2  ({\cal N}+u))^4
   (1 + 2  ({\cal N}+u))^2 +\\
+ 768  (\alpha  \epsilon^2 - 2  \delta  \epsilon  \gamma + 4
\gamma^2  \zeta)^2 +\\ +
  32 \gamma^4 (-1 + 2 ({\cal N}+u))^2 (-1 - 12 ({\cal N}+u) + 12 ({\cal N}+u)^2) \cdot \\
  \cdot  (3 \alpha^2 \epsilon^2 - 6 \alpha \delta \epsilon \gamma + 2 a \epsilon^2 \gamma + 2 \delta^2 \gamma^2 -
     4 d \epsilon \gamma^2 + 8 \gamma^3 z + 4 \alpha \gamma^2 \zeta) -\\
-
  256 \gamma^2(-1 + 2 ({\cal N}+u))^2 \cdot\\
\cdot (3 \alpha^2 \epsilon^3 - 9 \alpha \delta \epsilon^2 \gamma +
     a \epsilon^3 \gamma + 6 \delta^2 \epsilon \gamma^2 - 3 d \epsilon^2 \gamma^2 + 2 \delta^2 \gamma^4 + \\
+
     2 d \epsilon \gamma^4 + 12 \epsilon \gamma^3 z
- 4 \gamma^5 z + 12 \alpha \epsilon \gamma^2 \zeta -
     12 \delta \gamma^3 \zeta + 4 \alpha \gamma^4 \zeta)
 \end{array}
\label{eq:Phi1}
\end{equation}
\item[{\bf Case 2}:]  $\gamma = 0, \; \epsilon \ne 0 $\\
$$\rho({\cal N}) =1$$
\begin{equation}
\begin{array}{l}
\Phi( {\cal N} )=\\ = \frac{1}{4}
 \left( -\frac{K}{\epsilon} - \frac{z}{\sqrt{\epsilon}} - \frac{\delta}{\sqrt{\epsilon}}   \frac{\zeta}{\epsilon} +
\frac{\zeta^2}{\epsilon^2} \right) -\\ -\frac{1}{12}\Big( 3 d  - a
\sqrt{\epsilon} - 3 \alpha  \frac{\delta}{\sqrt{\epsilon}} + 3
\left(\frac{\delta}{\sqrt{\epsilon}}\right)^2- 6
\frac{z}{\sqrt{\epsilon}} +6 \alpha \frac{\zeta}{\epsilon}- 6
\frac{\delta}{\sqrt{\epsilon}}   \frac{\zeta}{\epsilon}\Big)
({\cal N}+u)\\ + \frac{1}{4} \left( \alpha^2 + d - a
\sqrt{\epsilon} - 3 \alpha \frac{\delta}{\sqrt{\epsilon}}+ \left(
\frac{\delta}{\sqrt{\epsilon}}\right)^2+ 2 \alpha
\frac{\zeta}{\epsilon} \right)  ({\cal N}+u)^2-\\ -\frac{1}{6}
\left( 3 \alpha^2 - a \sqrt{\epsilon} - 3 \alpha
\frac{\delta}{\sqrt{\epsilon}} \right) ({\cal N}+u)^3 +\frac{1}{4}
\alpha^2 ({\cal N}+u)^4
\end{array}
\label{eq:Phi2}
\end{equation}
\end{itemize}

The above formula is valid for $\epsilon>0$.

\section{Finite dimensional representations of qua\-dratic algebras}
\label{sec:Representations}
 Let consider a representation of the
quadratic algebra , which is diagonal to the generator $A$ and the
Casimir $K$. Using the parafermionic realization defined by
equations (\ref{eq:A}) and (\ref{eq:B}), we see that this a
representation diagonal to the parafermionic number operator
${\cal N}$ and the Casimir $K$. The basis of a such representation
corresponds to the Fock basis of the parafermionic oscillator,
i.e. the vectors $|k, \, n >,\; n=0,1,\ldots $of the carrier Fock
space satisfy the equations $$ {\cal N}   |k, \, n >= n|k, \, n >,
\quad K|k, \, n
>= k |k, \, n
> $$ The structure function (\ref{eq:Phi1}) (or respectively
(\ref{eq:Phi1}) ) depend on the eigenvalues of the of the
parafermionic number operator ${\cal N}$ and the  Casimir $K$. The
vectors $|k, \, n >$ are also eigenvectors of the generator $A$,
i.e. $$ A  |k, \, n >= A(k,n)|k, \, n > $$

In the case $\gamma \ne 0$ we find from equation (\ref{eq:sol1})
$$ A\left(k,n\right) =
 \frac{\gamma}{2} \left(
 (n+u)^2-1/4
-\frac{\epsilon}{ \gamma^2} \right) $$

In the case $\gamma = 0,\; \epsilon \ne 0$ we find from equation
(\ref{eq:sol1a})
 \[ A(k,n) = \sqrt{\epsilon}  \left( n+u \right).
\]

If the deformed oscillator corresponds to a deformed Parafermionic
oscillator of order $p$ then the two parameters of the calculation
$k$  and $u$ should satisfy the constrints (\ref{eq:restriction1})
and (\ref{eq:restriction2}) the system:
\begin{equation}
\begin{array}{c}
\Phi(0,u,k)=0 \\ \Phi(p+1,u,k)=0
\end{array}
\label{eq:system}
\end{equation}
then the parameter $u=u(k,p)$ is a solution of the system of
equations (\ref{eq:system}).

Generally there are many solutions of the above system, but a
unitary representation of the deformed parafermionic oscillator is
restrained by the additional restriction
\[ \Phi(x) >0, \quad
\mbox{for}  \quad x=1,2,\ldots,p \]
 We must point out that the system
(\ref{eq:system}) corresponds to a  representation with dimension
equal to $p+1$.

The proposed method of calculation of the representation of the
quadratic algebra is an alternative to the method given by
Granovskii et al \cite{GLZ-1992,GLZ-R1,GLZ-R2,GLZ-R3} and reduces
the search of the representations to the solution of a system of
polynomial equations (\ref{eq:system}). Also its is applied to an
algebra not included in the cases of the algebras, which are
treated in the above references. We must point out, that there are
several papers on the representations of quadratic ( or generally
polynomial algebras)
\cite{Karassiov94,Karassiov98,BoKolDas95,AbBeChaDe96,Debergh98,BeBriDe99,VanJag95},
these algebras are deformations of the su(2) or osp(1/2) algebras.
 The general form of the quadratic algebra,
which is studied in this paper, is different by definition from
the deformed versions of su(2) or osp(1/2).

\section{Quadratic algebras for the quantum superintegrable systems}
\label{sec:QuantumQuadratic}
 In this section, we shall give an
example of the calculation of eigenvalues of a superintegrable
two-dimensional system, by using the methods of the previous
section. The calculation by an empirical method was performed in
\cite{BoDasKo94} and the solution of the same problem by using
separation of variables was studied in \cite{KaMiPogo96}.  Here in
order to show the effects of the quantization procedure we don't
use $\hbar=1$ as it was considered in references  \cite{BoDasKo94}
and  \cite{KaMiPogo96}. That means that the following commutation
relations are taken in consideration:
\[
[x,p_x]= i \hbar, \qquad [y,p_y]= i \hbar
\]

\subsection{Potential i)}
 \[\displaystyle H=
\frac{1}{2}\left(p_x^2+p_y^2+ \omega^2 r^2+
\frac{\mu_1}{x^2}+\frac{\mu_2}{y^2}\right)
\]
This potential has the following independent integrals of motion:
\[
A=p_x^2+\omega^2 x^2+\frac{\mu_1}{x^2}
\]
and
\[
 B=\left(x p_y - y
p_x\right)^2+r^2\left(\frac{\mu_1}{x^2}+\frac{\mu_2}{y^2}\right)
\]
The constants, which characterize  the corresponding quadratic
algebra (\ref{eq:prealgebra}), are given by:
\[
\begin{array}{lll}
\alpha= 8\hbar^2,& \gamma=0,& \delta=-16 h^2 H,\\
 \epsilon=16\hbar^2 \omega^2,&
\zeta=-16 \hbar^2(\mu_1+\mu_2)\omega^2+8\hbar^4\omega^2\\ a=0,&
d=16 \hbar^4, & z=-16 \hbar^2(\mu_2-\mu_1)\omega^2-16 \hbar^4 H
\end{array}
\]
the value of the Casimir (\ref{eq:Casimir}) is:
\[
K=16 \hbar^2 \left((\mu_2-\mu_1)^2\omega^2+4 \mu_1 H^2 \right) -16
\hbar^4 \left(3 H^2 +2\hbar^2 \omega^2 -2(\mu_1+\mu_2)\right)
\]
For simplicity reasons we introduce the positive parameters $k_1$
and $k_1$, which are related to the potential parameters $\mu_1$
and $\mu_2$ by the relations:
\[
\mu_1 = \left(k_1^2- \frac{1}{4}\right)\hbar^2 \qquad \mu_2 =
\left(k_2^2- \frac{1}{4}\right)\hbar^2
\]

 This quadratic algebra corresponds to the case $\gamma=0$ and
$\epsilon >0$ of the algebra given by equations
(\ref{eq:algebra1}--\ref{eq:algebra3}). In that case, the
structure function (\ref{eq:Phi2}) of the deformed parafermionic
algebra of Section \ref{sec:Para} can be given by the simple form:
\[
\begin{array}{rl}
\Phi(x)=& 16\hbar^4 \left( x+u -\frac{1}{2}-\frac{k_1}{2}\right)
\left( x+u -\frac{1}{2}+\frac{k_1}{2}\right)\cdot \nonumber\\
 &\cdot \left( x+u
-\frac{1}{2}-\frac{k_2}{2}-\frac{E}{2\hbar\omega}\right) \left(
x+u -\frac{1}{2}+\frac{k_2}{2}-\frac{E}{2\hbar\omega}\right)
\end{array}
\]
In the above formula $E$ is the eigenvalue of the energy. The
values of the parameters $u$ and $E$ corresponding to the a
representation of the parafermionic algebra of dimension $p+1$ are
determined by the restrictions (\ref{eq:system}), which are
transcribed as:
\[
\Phi(0)=0,\qquad \Phi(p+1)=0
\]
One should notice, that only the solutions which correspond to
positive eigenvalues of the integral $A$ must be retained. The
acceptable solutions are four and correspond to the following
values of the parameters $u$ and $E$:
\[
u=\frac{1}{2} +\frac{\epsilon_1 k_1}{2}, \qquad
 E=2 \hbar \omega
\left( p+1 + \frac{\epsilon_1 k_1+\epsilon_2 k_2}{2}\right)
\]
where $\epsilon_i=\pm 1$. The corresponding structure function is
\[
\Phi(x)= 16\hbar^4 x \left(p+1-x\right)
 \left( x+{\epsilon_1 k_1}\right)
 \left(p+1-x+{\epsilon_2 k_2}\right)
\]
The corresponding eigenvalues of the operator $A$ are given by:
\[
A(m)= 4 \hbar \omega \left( m +  \frac{\epsilon_1 k_1+\epsilon_2
k_2}{2}\right), \qquad m=0,1,\ldots,p
\]
The structure function $\Phi(x)$ should be a positive function,
for $x=1,2,\ldots,p$ therefore the constants $k_i$ are restricted
by the relations:
\[
\epsilon_1 k_1 > -1, \qquad \epsilon_2 k_2 > -1
\]

\subsection{Potential ii)}
\[
\displaystyle H =\frac{1}{2}\left(p_x^2+p_y^2+
  \omega^2\left(4x^2+y^2\right)+
\frac{\mu}{y^2}\right)
\]
This potential has the following independent integrals of motion:
\[
A=p_x^2+4\omega^2 x^2
\]
and
\[
 B=\frac{1}{2}\left\{x p_y -y p_x,p_y \right\}+ \frac{\mu x}{y^2} -\omega^2 x y^2
\]
The constants, which characterize  the corresponding quadratic
algebra (\ref{eq:prealgebra}), are given by:
\[
\begin{array}{lll}
\alpha=0,&\gamma=0,& \delta=0,\\
 \epsilon=16 \hbar^2 \omega^2,&
\zeta=0,\\ a=6 \hbar^2, & d=-16\hbar^2 H,&z=-8\hbar^2
(\mu\omega^2- H^2)+6 \hbar^4 \omega^2
\end{array}
\]
the value of the Casimir (\ref{eq:Casimir}) is:
\[
K=64 \hbar^4 \omega^2 H
\]
For simplicity reasons we introduce the positive parameters $k$,
which is related to the potential parameter $\mu$ by the relation:
\[
\mu = \left(k^2- \frac{1}{4}\right) \hbar^2
\]

 This quadratic algebra corresponds to the case $\gamma=0$ and
$\epsilon >0$ of the algebra given by equations
(\ref{eq:algebra1}--\ref{eq:algebra3}). In that case, the
structure function (\ref{eq:Phi2}) of the deformed parafermionic
algebra of Section \ref{sec:Para} can be given by the simple form:
\[
\Phi(x)=8 \hbar^3 \omega \left(x+u-\frac{1}{2}\right)
\left(x+u-\frac{1}{2}-\frac{k}{2}-\frac{E}{2\hbar\omega} \right)
\left(x+u-\frac{1}{2}+\frac{k}{2}-\frac{E}{2\hbar\omega} \right)
\]
In the above formula $E$ is the eigenvalue of the energy. The
values of the parameters $u$ and $E$ corresponding to the a
representation of the parafermionic algebra of dimension $p+1$ are
determined by the restrictions (\ref{eq:system}), which are
transcribed as:
\[
\Phi(0)=0,\qquad \Phi(p+1)=0
\]
One should notice, that only the solutions which correspond to
positive eigenvalues of the integral $A$ must be retained. The
acceptable solutions are four and correspond to the following
values of the parameters $u$ and $E$:
\[
u=\frac{1}{2} , \qquad E=2 \hbar \omega \left( p+1 +
\frac{\epsilon k}{2\hbar}\right)
\]
where $\epsilon=\pm 1$. The corresponding structure function is
\[
\Phi(x)= 4\hbar^3 x \left(p+1-x\right)
 \left( p+1-x+{\epsilon k}\right)
\]
The structure function should be a positive function, therefore
the values of the parameter $k$ are restrained by
\[
\epsilon k>-1
\]
The eigenvalues of the operator $A$ are given by:
\[
A(m)= 4 \hbar \omega ( m+ \frac{1}{2} ), \qquad m=0,1,\ldots,p
\]

\subsection{Potential iii)}
 \[\displaystyle H=
\frac{1}{2}\left(p_x^2+p_y^2
  +\frac{k}{r} + \frac{1}{r}\left(\frac{\mu_1}{r+x}+\frac{\mu_2}{r-x}\right) \right)
\]
In ref \cite{KaMiPogo96} the parabolic coordinates have been used:
\[
\begin{array}{ll}
x=\frac{1}{2}\left( \xi^2 - \eta^2\right), &
p_x=\frac{\xi}{\xi^2+\eta^2} p_\xi -\frac{\eta}{\xi^2+\eta^2}
p_\eta,\\
  y= \xi \eta,&
p_y=\frac{\eta}{\xi^2+\eta^2} p_\xi +\frac{\xi}{\xi^2+\eta^2}
p_\eta,\\ \left[ \xi, p_\xi \right]=i \hbar, & \left[ \eta, p_\eta
\right]=i \hbar
\end{array}
\]
For comparison reasons we quote all the relations in both,
cartesian and parabolic systems, so
\[
H=\frac{1}{\xi^2+\eta^2}\left(
 \frac{1}{2} \left( p_\xi^2+
p_\eta^2 \right) + k + \frac{\mu_1}{\xi^2} + \frac{\mu_2}{\eta^2}
\right)
\]
This potential has the following independent integrals of motion:
\[
\begin{array}{rl}
A=&(xp_y-yp_x)^2+r
\left(\frac{\mu_1}{r+x}+\frac{\mu_1}{r-x}\right)=\\ =& \frac{1}{2}
\left(\frac{1}{2}\left( \eta p_\xi - \xi p_\eta \right)^2 +
\left(\xi^2+\eta^2\right)\left(\frac{\mu_1}{\xi^2} +
\frac{\mu_2}{\eta^2}\right) \right)
\end{array}
\]
and
\[
\begin{array}{rl}
 B=&\frac{1}{2}\left(
 \left\{x p_y - y p_x, p_y\right\}-\frac{\mu_1}{
r}\frac{r-x}{r+x}+\frac{\mu_2}{ r}\frac{r+x}{r-x}+\frac{k x}{r}
\right)\\=& \frac{1}{\xi^2+\eta^2}\left( \frac{1}{2}\left( \xi^2
p_\eta^2- \eta^2 p_\xi^2\right) + \mu_2\frac{\xi^2}{\eta^2}-
 \mu_1\frac{\eta^2}{\xi^2}+\frac{k}{2} \frac{\xi^2-\eta^2}{\xi^2+\eta^2}
 \right)
\end{array}
\]
The constants, which characterize  the corresponding quadratic
algebra (\ref{eq:prealgebra}), are given by:
\[
\begin{array}{lll}
\alpha=0,&\gamma=2 \hbar^2,& \delta=0,\\
 \epsilon=-\hbar^4,& \zeta=- \hbar^2 k
(\mu_1-\mu_2), \\a=0,& d=8 \hbar^2 H,&z=-\hbar^2\left(4(\mu_1+
\mu_2)H- k^2/2\right)+\hbar^4 H
\end{array}
\]
the value of the Casimir (\ref{eq:Casimir}) is:
\[
K=  -\hbar^2\left( 2 ( \mu_1 - \mu_2)^2 H - k^2 ( \mu_1+ \mu_2)
\right)-  2 \hbar^4 \left( (\mu_1+ \mu_2)H- \frac{k^2}{4}\right)
+\hbar^6 H
\]
For simplicity reasons we introduce the positive parameters $k_1$
and $k_1$, which are related to the potential parameters $\mu_1$
and $\mu_2$ by the relations:
\[
\mu_1 =\frac{\hbar^2}{2} \left(k_1^2- \frac{1}{4}\right) \qquad
\mu_2 =\frac{\hbar^2}{2} \left(k_2^2- \frac{1}{4}\right)
\]

 This quadratic algebra corresponds to the case $\gamma \ne 0$
  of the algebra given by equations
(\ref{eq:algebra1}--\ref{eq:algebra3}). In that case, the
structure function (\ref{eq:Phi1}) of the deformed parafermionic
algebra of Section \ref{sec:Para} can be given by the simple form:
\[
\begin{array}{rrl}
\Phi(x)=& 3 \cdot 2^{14} \hbar^{16}& \left( 2 x -1 + k_1+k_2
\right) \left( 2 x -1 + k_1 - k_2 \right) \left( 2 x -1  - k_1 +
k_2 \right)\cdot
\\&\cdot & \left( 2 x -1 - k_1 - k_2 \right)  \left( 8 \hbar^2 H
x^2 - 8 \hbar^2 H x +2 \hbar^2 H +k^2 \right)
\end{array}
\]
In the above formula $E$ is the eigenvalue of the energy. The
values of the parameters $u$ and $E$ corresponding to the a
representation of the parafermionic algebra of dimension $p+1$ are
determined by the restrictions (\ref{eq:system}), which are
transcribed as:
\[
\Phi(0)=0,\qquad \Phi(p+1)=0
\]
One should notice, that only the solutions which correspond to
positive eigenvalues of the integral $A$ must be retained. The
acceptable solutions are four and correspond to the following
values of the parameters $u$ and $E$:
\[
u=\frac{1}{2}\left(2 +\epsilon_1 k_1 + \epsilon_2 k_2 \right),
\qquad E=-\frac{ k^2 }{ 2 \hbar^2 \left( 2(p+1) +\epsilon_1 k_1 +
\epsilon_2 k_2\right)^2}
\]
where $\epsilon_i=\pm 1$. The corresponding structure function is
\[
\begin{array}{rrl}
\Phi(x)=& 3\cdot 2^{20} \cdot k^2 \hbar^{16}&\cdot x (p+1-x)
\left(x + \epsilon_1 k_1 \right) \left(x + \epsilon_2 k_2
\right)\cdot \\ &\cdot&
 \left(x + \epsilon_1 k_1+\epsilon_2 k_2 \right)
 \frac{
\left(x +p+1+ \epsilon_1 k_1+\epsilon_2 k_2 \right)} { \left(
2(p+1) +\epsilon_1 k_1 + \epsilon_2 k_2\right)^2}
\end{array}
\]
The eigenvalues of the operator $A$ are given by the formula:
\[
A(m)= \hbar^2 \left( m +\epsilon_1 k_1+\epsilon_2 k_2 +
\frac{3}{2} \right)^2 , \qquad m=0,1,\ldots,p
\]
The positive sign  of the structure function for $x=1,2,\ldots,p$
is obtained when:
\[
\epsilon_1 k_1 > -1,\quad \epsilon_2 k_2 > -1,\quad \mbox{and}
\quad \epsilon_1 k_1+\epsilon_2 k_2 > -1
\]

\subsection{Potential iv)}

\[
\begin{array}{rl}
 H =&\frac{1}{2}\left(p_x^2+p_y^2+
  +\frac{k}{r} +
  \mu_1\frac{\sqrt{r+x}}{r}+\mu_2\frac{\sqrt{r-x}}{r
  } \right)=\\
=& \frac{1}{\xi^2+\eta^2} \left( \frac{1}{2}\left(
p_\xi^2+p_\eta^2\right)+k + \mu_1 \xi +\mu_2 \eta
 \right)
\end{array}
\]
This potential has the following independent integrals of motion:
\[
\begin{array}{rl}
A=&\frac{1}{2}\left(\left\{(yp_x-xp_y), p_y\right\}+ \frac{\mu_1
(r-x)\sqrt{r+u}}{r} -\frac{\mu_2 (r+x)\sqrt{r-u}}{r}-\frac{k x}{
r} \right)=\\ =& \frac{1}{2 \left( \xi^2+\eta^2\right)} \left(
\eta^2 p_\xi^2 - \xi^2 p_\eta^2  + k \left(\eta^2 - \xi^2 \right)
+ 2 \xi \eta \left( \mu_1 \eta - \mu_2 \xi\right) \right)
\end{array}
\]
and
\[
\begin{array}{rl}
 B=&\frac{1}{2} \left(\left\{x p_y - y p_x,p_x \right\}- \frac{\mu_1 x\sqrt{r-u}}{r}
+\frac{\mu_2 x\sqrt{r+u}}{r}-\frac{k y}{ r}\right)=\\ =&
-\frac{1}{2 \left( \xi^2+\eta^2\right)} \left( \xi \eta \left(
p_\xi^2 + p_\eta^2 \right)- \left( \xi^2+\eta^2\right)p_\xi p_\eta
+2 k \xi \eta +\left( \mu_2 \xi - \mu_1 \eta \right)\left( \eta^2
- \xi^2 \right)
 \right)
\end{array}
\]
The constants, which characterize  the corresponding quadratic
algebra (\ref{eq:prealgebra}), are given by:
\[
\begin{array}{lll}
\alpha=0,& \gamma=0,& \delta=0,\\
 \epsilon=-2\hbar^2 H,& \zeta= \hbar^2 \mu_1
\mu_2/2 ,\\ a=0,& d= 2 \hbar^2 H,& z=- \hbar^2
{(\mu_1^2-\mu_2^2)}/{4}
\end{array}
\]
the value of the Casimir (\ref{eq:Casimir}) is:
\[
K=\hbar^2 {k^2}H/{2}  + \hbar^2 k(\mu_1^2+\mu_2^2)/{4}+ \hbar^4
H^2
\]

 This quadratic algebra corresponds to the case $\gamma=0$ and
$\epsilon >0$ of the algebra given by equations
(\ref{eq:algebra1}--\ref{eq:algebra3}). It is worth noticing that
the algebra is extremely simple, which can be reduced to the usual
$su(2)$. We prefer to treat this algebra with the proposed methods
in this paper for pedagogical reasons. The existence of the finite
dimensional representations of this algebra implies that, the
coefficient $\epsilon$ in equation (\ref{eq:algebra2}) should be
positive, therefore the energy operator $H$ must have energy
eigenvalues $E<0$. For simplicity reasons we introduce the new
parameters:
\[
\begin{array}{lll}
\varepsilon= \sqrt{-2 E}/\hbar,& \lambda= k/\hbar^2,\\ \nu_1=
\mu_1/\hbar^2, & \nu_2= \mu_2/\hbar^2,& \nu^2=\nu_1^2 + \nu_2^2
\end{array}
\]

The structure function (\ref{eq:Phi2}) of the deformed
parafermionic algebra of Section \ref{sec:Para} can be given by
the form:
\[
\Phi(x)=\frac{\hbar^4}{16\varepsilon ^4} \left( \nu_1^2 - \lambda
\varepsilon^2+2 (x+u-\frac{1}{2}) \varepsilon^3\right) \left(
\nu_2^2 - \lambda \varepsilon^2-2 (x+u-\frac{1}{2})
\varepsilon^3\right)
\]
In the above formula the parameter $\varepsilon$ is related to the
  the eigenvalue $E$ of the energy. The values of the parameters
$u$ and $\varepsilon$, corresponding to the a representation of
the parafermionic algebra of dimension $p+1$, are determined by
the restrictions (\ref{eq:system}), which are transcribed as:
\[
\Phi(0)=0,\qquad \Phi(p+1)=0
\]
The first condition can be used for determining the acceptable
values of the parameter $u$. Two possible solutions are found:
\begin{eqnarray}
u=u_1= \frac{ \nu_2^2 -\lambda \varepsilon^2 + \varepsilon^3}{2
\varepsilon^3} \label{eq:Pot4u1}\\
 u=u_2= -\frac{ \nu_1^2 -\lambda
\varepsilon^2 - \varepsilon^3}{2 \varepsilon^3} \label{eq:Pot4u2}
\end{eqnarray}
Using these solutions and the condition $Phi(p+1)=0$, we find that
the $\varepsilon$ must satisfy two possible cubic equations:
\begin{eqnarray}
u_1\; \longrightarrow\; 2(p+1) \varepsilon^3 - 2 \lambda
\varepsilon^2 + \nu^2=0 \label{eq:Pot4e1}\\ u_2\;
\longrightarrow\; 2(p+1) \varepsilon^3 + 2 \lambda \varepsilon^2 -
\nu^2=0 \label{eq:Pot4e2}
\end{eqnarray}
If $\varepsilon$ is a solution of equation (\ref{eq:Pot4e1}) then
$-\varepsilon$ is the solution of the other equation
(\ref{eq:Pot4e2}), therefore there is almost one solution which is
positive. This solution leads to the structure function:
\[
\Phi(x)= \frac{\varepsilon^2}{4} x \left( p+1-x\right)
\]
which is positive for $x=1,2,\ldots,p$.

\section{Discussion}\label{sec:Discussion}

If we compare the quadratic associative algebra, introduced in
section \ref{sec:AssocQuadrAlgebra} with the corresponding Poisson
algebra given in section \ref{sec:Poisson}, we see that in general
the quantum constants are  similar to the classical ones up to an
factor equal to $-h^2$, but there are quantum corrections of order
$h^4$ and $h^6$. The knowledge of the classical constants of the
Poisson algebra is not sufficient to reproduce the rules of
quantum  associative operator algebra.  Therefore, the passage
from the classic Poisson algebra to the non commutative quantum
algebra can not be realized by simple replacements of the Poisson
brackets by commutators and by a symmetrization procedure.

 The energy eigenvalues of section \ref{sec:QuantumQuadratic} corroborate
  the results of reference \cite{KaMiPogo96}
(the differences in the case of the potential iv are due to some
misprints in that reference). The calculation of the energy
eigenvalues in  reference \cite{KaMiPogo96} was achieved by
solving the corresponding Schroedinger differential equations,
while in this paper the energy eigenvalues are obtained by
algebraic methods. The advantage of the proposed method is that,
the energy eigenvalues are reduced to simple algebraic
calculations of the roots of polynomial equations, whose the form
is universally determined by the the structure functions
(\ref{eq:Phi1}), (\ref{eq:Phi1}) and the system (\ref{eq:system}).
These equations are valid for any two dimensional superintegrable
system with integrals of motion, which are quadratic functions of
the momenta. The same equations should be valid in the case of two
dimensional superintegrable systems in curved space
\cite{Istanbul00}. The superintegrable systems bring up for
discussion the open problem of the quantization of a Poisson
algebra in a well determined context, because these systems and
their quantum counterparts are explicitly known.

From the above discussion several open problems are risen:
\begin{itemize}
\item
The calculation of the classical Poisson algebras and their
quantum counterparts for the totality of the two dimensional
problems in curved space. This study will lead to the calculation
of the energy eigenvalues by algebraic methods.
\item
The two dimensional superintegrable systems are classified by the
values of the constants of the Darboux conditions \cite{RaSan99}.
The relation of these constants with the constants of the
quadratic Poisson algebra is not yet known.
\item
The Poisson algebras  for the Drach superintegrable systems with a
cubic integral of motion were written by using a classical
analogue of the  deformed parafermionic algebra \cite{Tsiganov00}.
Their quantum counterparts and the calculation of their energy
eigenvalues are is a topic under investigation.
\item
The Poisson algebras and the associated quantum counterparts for
the three dimensional superintegrable systems are not yet fully
studied. Recently \cite{KaWiMiPo99} the quantum quadratic algebras
have been written down, but  a systematic calculation of energy
eigenvalues was not yet performed.
\end{itemize}

The above points show that, the study of non linear Poisson
algebras and their quantum counterparts is a  topic of interest.

\end{document}